\documentclass[aps,prl,twocolumn,showpacs]{revtex4}%
\usepackage{graphicx}
\usepackage{amsmath}
\usepackage{amsfonts}
\usepackage{amssymb}%
\newcommand{\be}{\begin{equation}}
\newcommand{\ee}{\end{equation}}
\newcommand{\bea}{\begin{eqnarray}}
\newcommand{\eea}{\end{eqnarray}}

\begin{document}
\title{Effect of spin-orbit interaction on a magnetic impurity in the vicinity of a surface}
\author{L. Szunyogh,$^{1,2}$ G. Zar\'and,$^{1}$ S. Gallego,$^{3}$ M.~C. Mu\~noz,$^{3}$ and B.~L.  Gy\"orffy$^{2,4}$ }
\affiliation{$^{1}$~Department of Theoretical Physics and Center for Applied Mathematics
and Computational Physics, \\
Budapest University of Technology and Economics, Budafoki \'ut. 8, 1521
Budapest, Hungary \\ 
$^{2}$~Center for Computational Materials Science, TU Vienna, Getreidemarkt
9/134, A-1060 Vienna, Austria \\ 
$^{3}$~Instituto de Ciencia de Materiales de Madrid, Consejo Superior de
Investigaciones Cient{\'{\i}}ficas, Cantoblanco, 28049 Madrid, Spain \\
$^{4}$~H.H. Wills Physics Laboratory, Bristol University, Royal Fort, Tyndall
Avenue, Bristol BS8 1TL, U.K. }
\date{\today}

\begin{abstract}
We propose a new mechanism for surface-induced magnetic anisotropy to explain
the thickness-dependence of the Kondo resistivity of thin films of dilute
magnetic alloys. The surface anisotropy energy, generated by spin-orbit
coupling \emph{on the magnetic impurity itself}, is an oscillating function of
the distance $d$ from the surface and decays as 1/$d^{2}$. Numerical estimates
based on simple models suggest that this mechanism, unlike its alternatives,
gives rise to an effect of the desired order of magnitude.

\end{abstract}

\maketitle



The original observation that the amplitude 
of the Kondo resistivity
in thin films of dilute magnetic alloys depends on the thickness 
$L$ of the film, \cite{chen,ditusa} has attracted considerable attention
\cite{blachly,giordano,jacobs,orsi1,orsi2}. So far the most promising
explanation of this phenomenon was given by  \'{U}js\'{a}ghy \emph{et al.}
\cite{orsi1,orsi2}: They suggested that a magnetic impurity, such as Fe in Au,
near the surface of the host metal is subject to a magnetic anisotropy that
blocks the spin--dynamics responsible for the Kondo scattering within a given
distance, $L_{c}$, to the surface, if the anisotropy is larger 
than the Kondo temperature $T_K$ of the magnetic impurities.
As a consequence, the Kondo
resistivity $\ R_{K}(L)$ of a thin film relative to that of a 'bulk--like'
thick film, 
$R^{\rm bulk}_{K}(L)$, can be estimated for $\ L\gg L_{c}$ as
\begin{equation}
\frac{R_{K}(L)}{R^{\rm bulk}_{K}(L)}=1-\frac{2L_{c}}{L}\; ,
\end{equation}
where the factor 2 on the \emph{rhs} accounts for the two surfaces of the
film. Fitting the experiments with this formula yields $L_{c}\simeq180$
\AA \ in case of Au(Fe) thin films \cite{orsi1}.

\'{U}js\'{a}ghy et al. \cite{orsi1,orsi2} based their arguments on
calculations for 
a 5/2 Kondo impurity embedded in a
semi-infinite host with spin-orbit interaction on the host atoms only. They
found a surface induced anisotropy energy,
\[
H_{\mathrm{anis}}=K(d)\widehat{S}_{z}^{2}\;,
\]
where $\widehat{S}_{z}$ is the z-component of the impurity spin operator and
the anisotropy constant $K(d)$ is a function of the distance $d$
of the impurity from the surface. Using lowest order perturbation theory in
both the spin-orbit coupling constant $\lambda$ and the effects of the surface
they derived an expression for $K(d)$ which for large $d$ decayed as $1/d$.
Moreover, their quantitative estimates of $K(d)$ gave sufficiently large
values to explain the experimental facts in their more detailed calculations
\cite{orsi3}.

Subsequently, Szunyogh and Gy\"orffy~\cite{SG:prl97} studied the problem using
a material specific, parameter-free first-principles approach, namely,
spin-polarized relativistic calculations based on the Local Density
Approximation (LDA). In these calculations, the semi-infinite host was taken 
into account without significant approximations. Consequently, the spin-orbit
coupling on the host atoms and electrons' scattering between the impurity and
the host atoms were treated on equal footing and to all orders in the coupling
strength. They found $K(d)\approx\cos(2k_{F}^{\prime}d)/d^{2}$, where
$2k_{F}^{\prime}$ is the length of a spanning vector of the host's Fermi
Surface. 
Moreover, the size of the anisotropy energy turned out to be too
small by orders of magnitude. Clearly, the LDA results would eliminate the whole 
idea of explaining the size dependent Kondo effect by 'surface induced magnetic anisotropy', 
if they didn't suffer from the well-known weakness of the LDA in describing
spin-fluctuations on the impurity. 
LDA calculations also underestimate 'Hunds rule
correlations' systematically.  

\begin{figure}[ptb]
\begin{center}
\includegraphics[width=0.45\textwidth,clip,bb=25 135 530 460]{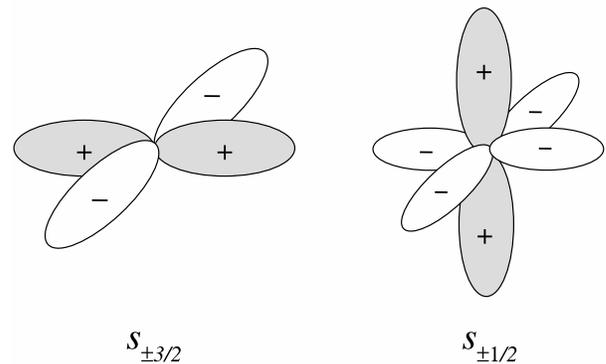}
\end{center}
\par
\vspace{-0.60cm}\caption{Sketch of the orbital structures of the $d$--type
$\Gamma_{8}$ combinations formed by $s$--orbitals at the nearest--neighbor
sites around an impurity in a simple cubic host. The $s_{\pm3/2}$ combinations
are composed from orbitals located at neighbors within the same plane, whereas
the $s_{\pm1/2}$ combinations have significant out-of-plane contributions. }%
\label{fig:d-orbitals}%
\end{figure}

In this Letter we present calculations  
which include dynamical spin-fluctuations at the same
level as Refs.~\cite{orsi1,orsi2}, but go beyond their approach 
in that here we treat the spin-orbit coupling in 
the semi-infinite host non-perturbatively, and we also incorporate 
the effects of spin-orbit coupling on the impurity.
We find that, while the host-induced anisotropy, as proposed in 
Ref.~\cite{orsi1}, is negligible, an improved treatment of 
correlations and the strong (typically $\sim1$ eV) spin-orbit coupling {\em on  the magnetic impurity}
can lead to a dramatic enhancement of the surface induced magnetic anisotropy
for impurities with partially filled $d$-shell.  As it turns out
it is large enough to explain the experiments. 
Furthermore, this anisotropy  has a simple physical origin: for partially filled
$d$-shells each spin-state has also an \emph{orbital structure}. In a given
spin-state, electrons on the deep $d$-levels lower their energy by hybridizing
with the conduction electrons through virtual fluctuations. However, in the
vicinity of a surface Friedel oscillations appear and, therefore, the density
of states available for the $d$-electrons to hybridize with depends on the
\emph{orbital state}, and, thus, on the spin of the impurity (see
Fig.~\ref{fig:d-orbitals}). This mechanism gives rise to an anisotropy that
already appears to first order of the exchange coupling $J$ between the
magnetic impurity and the conduction electrons, decays as $1/d^{2}$ and is
orders of magnitude larger than the similar anisotropy induced by spin-orbit
coupling on the host sites \cite{orsi1}.

We shall first analyze the simplest possible model that captures
all important aspects of the problem, and  start with a
magnetic impurity in a $d^{1}$ configuration such as a $V^{4+}$ or $Ti^{3+}$
ion embedded into a simple cubic (sc) lattice. In this case, by Hund's third
rule, strong local spin-orbit coupling will lead to a $J=3/2$ multiplet that
is separated from the $J=5/2$ state typically by an energy of the order of $\sim1$ eV.
The advantage of this model is that under a cubic crystal field the $J=3/2$
multiplet remains degenerate ($\Gamma_{8}$ double representation), and no
anisotropy is generated when the surface is absent. For the sake of
simplicity, we assume that the host atoms form a (001) surface of a simple
cubic lattice. We also assume that the conduction band of the host is
dominated by $s$-electrons, which we describe in terms of a single--band
nearest--neighbor tight-binding model. 

The impurity's $J=3/2$ multiplet can only hybridize with those linear
combinations of $s$-states on the nearest neighbor atoms which transform
according to the $\Gamma_{8}$ representation. Within the full 12--dimensional
subspace, spanned by orbitals on six neighbors and two spin-states per site,
two such (four--dimensional) sets can be constructed. One of these has
$p$-type orbital structure and, therefore, hybridizes only weakly with the
impurity's $d$-level. The $d$-type set, $s_{m} $ ($m=-3/2,\dots,3/2$), which
hybridizes strongly with the $J=3/2$ multiplet, splits into combinations with
two different orbital characters as depicted in Fig.~\ref{fig:d-orbitals}: the
combinations $s_{\pm1/2}$ couple to the impurity states $J_{z}=m=\pm1/2$,
while $s_{\pm3/2}$ couple to $J_{z}=m=\pm3/2$.

Assuming that the impurity--host interaction is mainly dominated by quantum
fluctuations to the (non--degenerate) $d^{0}$ state, in lowest order of the
hybridization, a Coqblin--Schrieffer transformation leads to the following
effective exchange interaction, \cite{coqblin}
\begin{equation}
H_{J}=J\sum_{m,m^{\prime}}s_{m}^{\dagger}s_{m^{\prime}} \: \mid\! \frac{
\mbox{\footnotesize{3}} }{ \mbox{\footnotesize{2}} } m^{\prime}\rangle
\langle\frac{ \mbox{\footnotesize{3}} }{ \mbox{\footnotesize{2}} } m \!
\mid\;,
\end{equation}
where $|\frac{3}{2}m\rangle$ stand for the four states of the $\Gamma_{8}$
impurity multiplet, $s_{m}^{\dagger}$ and $s_{m}$ are creation and
annihilation operators acting on the host states, respectively, and $J$
denotes the effective strength of the coupling. We then employed Abrikosov's
pseudo-fermion representation \cite{pseudofermion} to calculate the splitting
of the four states up to second order in $J$ \cite{orsi1}. The first and
second order contributions to the self energy
at $T=0$ are given by
\begin{align}
&  \Sigma_{mm^{\prime}}^{(1)}=J\int_{-\infty}^{E_{F}}d\epsilon\varrho
_{mm^{\prime}}(\epsilon)\;,\label{eq:sigma1}\\
\ \hspace{-0.6cm} {\rm and}  \quad & \ \nonumber\\
&  \Sigma_{mm^{\prime}}^{(2)}=J^{2}\sum_{m^{\prime\prime}}\int_{-\infty
}^{E_{F}}d\epsilon\int_{E_{F}}^{\infty}d\epsilon^{\prime}\nonumber\\
&  \qquad\qquad\qquad\qquad{\frac{1}{\epsilon^{\prime}-\epsilon}}%
\varrho_{mm^{\prime}}(\epsilon)\varrho_{m^{\prime\prime}m^{\prime\prime}%
}(\epsilon^{\prime})\;, \label{eq:sigma2}%
\end{align}
respectively. Here $\varrho_{mm^{\prime}}(\epsilon)$ denotes the local
spectral function of the host computed in the absence of the exchange
interaction, \emph{i.e.}, $J=0$, and $E_{F}$ is the Fermi 
energy.~\cite{cut-off} To compute
$\varrho_{mm^{\prime}}(\epsilon)$, we made use of the so--called surface
Green's function matching procedure~\cite{sgfm} that 
completely accounts for the semi--infinite geometry of the host.
Note that Eqs.~(\ref{eq:sigma1}) and (\ref{eq:sigma2}), in addition
to incorporating quantum fluctuations 
of the spin at the same level as Ref.~\cite{orsi1},
also take into account the semi-infinite nature 
of the host non-perturbatively through the spectral functions~\cite{host-soc}.

In this simple case, tetragonal symmetry of the sc(001) surface implies that
$\varrho_{m m^{\prime}}(\epsilon)$, consequently, $\Sigma_{m m^{\prime}}$ are
diagonal in $m, m^{\prime}$. From time reversal symmetry it further follows
that the states $|\frac{3}{2} \pm\!\! \frac{1}{2} \rangle$ and $|\frac{3}{2}
\pm\!\! \frac{3}{2} \rangle$ remain degenerate. Thus the fourfold degeneracy
of the $J=3/2$ multiplet is split by an effective anisotropy term,
\begin{equation}
H_{\mathrm{anis}} = K J_{z}^{2}\;,
\end{equation}
with $K = (\Sigma_{3/2} - \Sigma_{1/2})/2$.

\begin{figure}[ptb]
\begin{center}
\includegraphics[width=0.45\textwidth,clip,bb=56 429 488 755]{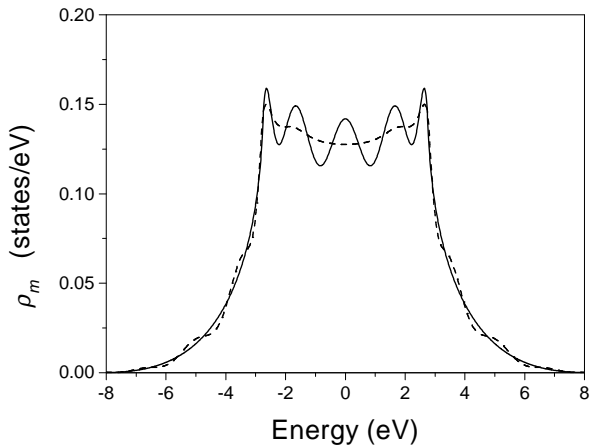}
\end{center}
\par
\vspace{-0.70cm}\caption{Spectral density functions corresponding to orbitals
$m=\pm\frac{3}{2}$ (solid line) and $m=\pm\frac{1}{2}$ (dashes), see also
Fig.~\ref{fig:d-orbitals}, at the fifth layer below the (001) surface of a sc
single--band host metal. The lattice parameter was chosen to 2.6 \AA \ and the
hopping to 1.4 eV. }%
\label{fig:d1-spdos}%
\end{figure}

A clear understanding of the level splitting due to the vicinity of a surface
obviously emerges from Fig.~\ref{fig:d-orbitals}: the spectral density related
to $s_{3/2}$ orbitals extending in a single atomic plane differs from that
corresponding to $s_{1/2}$ orbitals that in fact take an average over three
adjacent atomic planes. This is demonstrated in Fig.~\ref{fig:d1-spdos}, where
$\varrho_{\frac{3}{2}}(\epsilon)$ and $\varrho_{\frac{1}{2}}(\epsilon)$ are
plotted at the fifth atomic plane counted from the surface. In our numerical
implementation a layer spacing of 2.6 \AA \ (that corresponds to the atomic
volume of fcc Cu), and a hopping parameter $V=-1.4$ eV were chosen.
Evidently, $\varrho_{3/2}(\epsilon)$ oscillates strongly within the energy
range $-2|V|<\epsilon<2|V|$ (the center of the valence band is set to zero), while
$\varrho_{1/2}(\epsilon)$ behaves smoothly.

Departing from the surface, the oscillations of $\varrho_{3/2}(\epsilon)$ get
more and more rapid while decreasing in magnitude. For large $d$,
$\varrho_{1/2}(\epsilon)$ approaches $\varrho_{3/2}(\epsilon)$ \ and both tend
to $\varrho^{\mathrm{bulk}}(\epsilon)$. Interestingly, at a given energy, they
also display Friedel oscillations, \cite{lang-kohn}
\begin{equation}
\varrho_{m}(\epsilon;d)\simeq {A_{m}}\; \cos(2k_{z}(\epsilon)d+\phi)/d\;,
\label{eq:Friedel}%
\end{equation}
where $k_{z}(\epsilon)$ is the length of the extremal wave--vector parallel to
the surface normal of the constant--energy surface in reciprocal space and
$\phi=0$ or $\pi/2$. Performing the energy integrations in
Eqs.~(\ref{eq:sigma1}) and (\ref{eq:sigma2}), both self--energy contributions
yield oscillations with a period of $\pi/k_{z}(E_{F})$ for large $d$ and an
amplitude $\sim 1/d^{2}$, as also follows from an asymptotic
analysis analogous to Ref.~\onlinecite{SG:prl97}.

\begin{figure}[ptb]
\begin{center}
\includegraphics[width=0.45\textwidth,clip,bb=19 135 440 705]{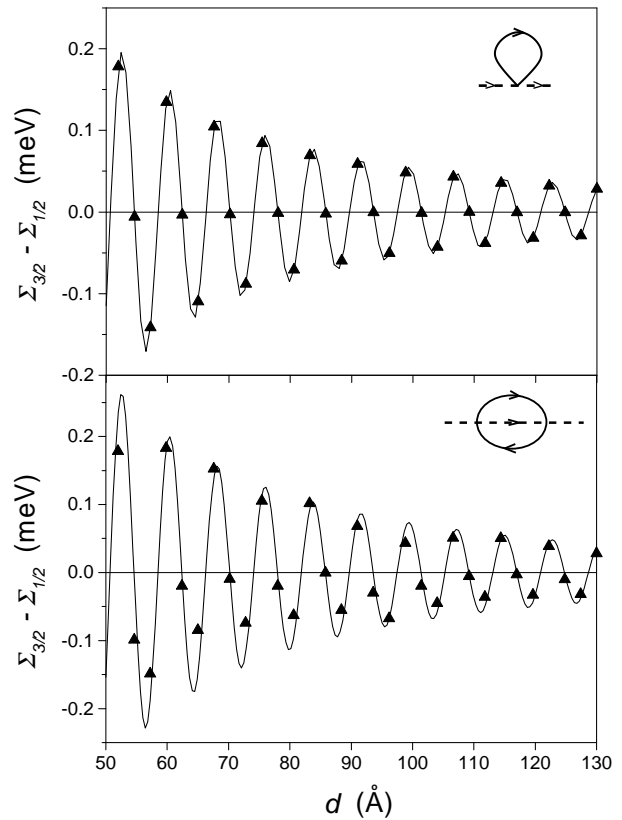}
\end{center}
\par
\vspace{-0.5cm}\caption{First order (upper panel) and second order
(lower panel) contributions to the level splitting of a $d^{1}$ impurity with
strong on--site spin--orbit interaction as a function of the distance $d$ from
a sc(001) surface. See for numerical parameters in the text. For both cases
the solid lines depict functions, $A \sin(2 \pi d/d_{0})/d^{2}$ ($d_{0}=$ 7.8
\AA ), fitted to the calculated values. }%
\label{fig:d1-anis}%
\end{figure}

The impurity's first and second order level splittings, $\Sigma_{3/2}%
-\Sigma_{1/2}$, are plotted in Fig.~\ref{fig:d1-anis} for $E_{F}=V=-$ 1.4 eV.
Here we used $J=1$ eV, a typical exchange coupling for
Kondo impurities with a Kondo temperature of the order of a few Kelvins.
As can also be obtained from analytic calculations, the period of the oscillations is 3 atomic layers
(7.8 \AA ) in this case. Remarkably, the second order self--energy diagram
contributes about the same amount to the level splitting as the first order 
one.

Note that the present theory also predicts that there will be 
impurities with nearly vanishing level splittings. However,
for incommensurate 
Friedel oscillations the distribution of  
$\Delta \equiv |\Sigma_{3/2}(d)- \Sigma_{1/2}(d)|$ 
within an interval $[d-\delta d, d + \delta d]$  of a few atomic layers  is {\em peaked} around the maximum values of 
the anisotropy within that interval,  $\Delta_{\rm max}(d)$. Therefore, 
for large values of $\Delta_{\rm max}(d) \gg T_K$ ({\em i.e.}, in the vicinity 
of the surface) only a very small fraction of the impurity spins will 
experience an anisotropy-induced splitting,  $\Delta < T_K$.
As can be
read off Fig.~\ref{fig:d1-anis}, beyond $d \sim 100$~\AA \ the amplitude 
$\Delta_{\rm max}$ is
still in the range of few tenths of a meV (1 K $\sim$ 0.09 meV), and
the typical 
level splitting is close to the values needed to suppress the Kondo effect in
thin films of alloys with $T_{K} \sim$ 0.1 -- 1 K such as Au(Fe).

As was stressed in the introduction,  the above mechanism is a combined
consequence of 'spin-orbit coupling' and 'Hunds rule' correlations at the magnetic 
impurity. Since the latter does not occur on the host sites we expect a 
much reduced contribution
to the total anisotropy energy from these. In fact we would expect that,
though the 'impurity spin' is treated classically, the host-induced anisotropy
is described reasonably well by the first-principles LDA calculations, hence,
the results for the anisotropy energy in Ref. \onlinecite{SG:prl97} can be
regarded as of the correct order of magnitude. 
This suggests that in models,
such as that of \'Ujs\'aghy et al.,~\cite{orsi1,orsi2,orsi3}
where the spin-orbit interaction is restricted to the host atoms, the 'surface
induced magnetic anisotropy' should be negligibly small. 
From this point of
view the fact that they find a level splitting sufficiently large to explain
the experimental data ought to be taken probably 
as a result of their largely
analytic approximations.  In fact, we also studied 
a somewhat modified version of our model where only host-induced 
spin-anisotropy appears and found, as in the first first principles 
calculations~\cite{SG:prl97}, that the obtained anisotropy was orders 
of magnitude smaller than the one discussed above.

\begin{figure}[ptb]
\begin{center}
\includegraphics[width=0.45\textwidth,clip,bb=25 405 445 720]{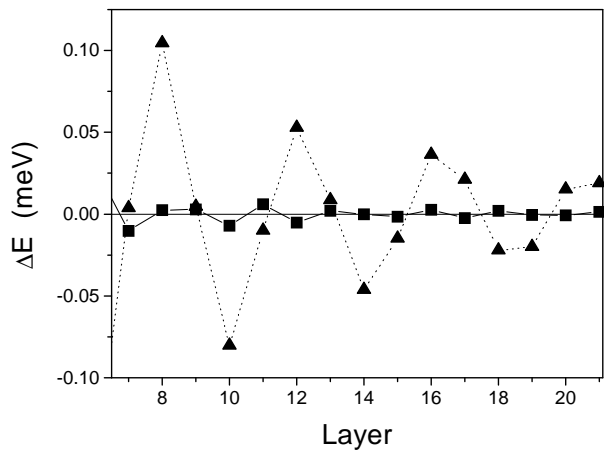}
\end{center}
\par
\vspace*{-0.50cm}\caption{ LDA magnetic anisotropy energies for an
Fe impurity in Au host as a function of it's layer position measured from the
(001) surface. Squares refer to the experimental lattice constant (7.68 \AA ),
while triangles to the case when this lattice constant has been uniformly
expanded by 50 \%.}%
\label{fig:feau}%
\end{figure}


To lend further credence to the new mechanism of surface induced anisotropy
we have attempted to simulate
the effects of Hunds rule correlations on the first principles calculations in
Ref.~\onlinecite{SG:prl97}. In calculations based on model Hamiltonians one
can manipulate the parameters of the theory, like the hopping integrals or the
impurity-electron interactions parameter, $J$, to desirable ends. Evidently,
there is no such freedom in first-principles calculations, however,
approximate. Nevertheless, in order to investigate correlations between
different effects one can constrain the outcome of such calculations in other
ways.
As a quite crude device, we increased the lattice spacing of the Au host to
mimic the band narrowing effects of correlations not included in LDA. In
Fig.~\ref{fig:feau} we compare the surface induced anisotropy energies of an
Fe impurity as calculated for the experimental lattice spacing with those for
a lattice constant artificially enhanced by 50 \%. Note that because of the
different lattice constants this comparison is made in terms of the impurity's
position measured in the layer index rather than in physical distances
from the surface. As expected from the considerations above, increasing
the lattice spacing resulted in a dramatic, order of magnitude increase of the
predicted anisotropy. Reassuringly, this result is quite robust since
it occurs in spite of the fact that, beside the band narrowing required to
increase the effect of the on-site spin-orbit coupling, such an increase of the
distance between the atoms should also reduce the coupling between the
impurity and the conduction electrons ($J$) and, hence, the anisotropy energy.
Clearly, the observed overall enhancement supports our contention that the
origin of the large anisotropy energy and long critical length, $L_{c}$, is,
indeed, the electron-electron correlations which are correctly captured by the
basic ansatz of our $d^{1}$ model but are neglected in the first principles
LDA calculations.

The authors are indebted to A. Zawadowski, O. \'Ujs\'aghy, and D.L. Cox for
valuable discussions. This work has been financed by the Hungarian National
Scientific Research Foundation (Contracts OTKA T046267 and T046303) and the
Ministerio de Ciencia y Tecnolog\'{\i}a (Spain) under contract No.
2004HU0010 and MAT 2003-04278. L.S. was also supported by the Center for 
Computational Materials Science (Contract No. GZ 45.531).


\begin{thebibliography}{99}
\bibitem{cut-off}
In reality, a cut-off around $E_F$
of the order of the spin-orbit coupling strength should appear in the
integrals,
however, this modification only influences our results at short distances, $d$.
\bibitem{host-soc}
Eqs.~(\ref{eq:sigma1}) and (\ref{eq:sigma2}) also 
apply in the presence of spin-orbit coupling in the host, which 
can be fully incorporated in the spectral densities, $\varrho_{mm^{\prime}}$.
\end{thebibliography}
\end{document}